\documentclass[times]{article}
\usepackage{isita2008}
\title{Secret Communication with Feedback}
\name{Deniz G\"{u}nd\"{u}z$^{\dagger}$$^{\ddagger}$, D. Richard Brown III$^{\dagger\dagger}$ and H. Vincent Poor$^{\dagger}$%
\thanks{This research was supported by the US National Science Foundation under grants  CCF-04-47743, ANI-03-38807, and CNS-06-25637.}
}
\address{
\begin{tabular}{cc}
\begin{tabular}{c}
$^{\dagger}$ Dept.~of Electrical Engineering, Princeton University, Princeton, NJ, 08544.\\
$^{\ddagger}$ Dept.~of Electrical Engineering, Stanford University, Stanford, CA, 94305.\\
$^{\dagger\dagger}$ Dept.~of Electrical and Computer Eng., Worcester Polytechnic Institute, Worcester, MA, 01609.
\end{tabular}
\end{tabular}
}

\usepackage[dvips]{hyperref}
\usepackage[dvips]{graphicx}
\usepackage[cmex10]{amsmath}
\usepackage{amsfonts}
\usepackage{amssymb}
\usepackage{amscd}
\usepackage{psfrag}

\newtheorem{thm}{Theorem}[section]
\newtheorem{cor}[thm]{Corollary}

\newtheorem{defn}{Definition}[section]

\begin{document}

\maketitle
\sloppy
\begin{abstract}
Secure communication with feedback is studied. An achievability scheme in which the backward channel is used to generate a shared secret key is proposed. The scenario of binary symmetric forward and backward channels is considered, and a combination of the proposed scheme and Maurer's coding scheme is shown to achieve improved secrecy rates. The scenario of a Gaussian channel with perfect output feedback is also analyzed and the Schalkwijk-Kailath coding scheme is shown to achieve the secrecy capacity for this channel.
\end{abstract}

\section{Introduction}

In his pioneering work \cite{Shannon:BSTJ:49}, Shannon introduced information theoretic security and defined perfect secrecy, which roughly refers to the case in which an enciphered cryptogram does not reveal any information to an eavesdropper about the underlying secret message. Shannon proved that perfect secrecy can be achieved with a shared secure key that is as long as the underlying message. Wyner showed in \cite{Wyner:BSTJ:75} that perfect secrecy can be achieved even without key distribution if the cryptogram is transmitted over a noisy broadcast channel in which the eavesdropper's channel is physically degraded with respect to the legitimate receiver's channel. This result was extended to more general broadcast channels in \cite{Csiszar:IT:78}, where it was shown that nonzero secrecy capacity can be achieved if the main channel is \emph{less noisy} than the eavesdropper's channel. Secure communication in the presence of eavesdroppers has gained a recent interest, and information theoretic security in various models has been explored in detail (see, for example, \cite{Tekin:IT:08}, \cite{Liang:IT:08a}, \cite{Khisti:IT:08} and \cite{Gunduz:ITW:08}).

While it is well-known that feedback doesn't increase the capacity of a point-to-point memoryless channel, it was observed in \cite{Cheong:Stanford:76} that the availability of feedback might increase the {\em secrecy capacity} of a point-to-point memoryless channel. This can be immediately seen by considering an infinite capacity secure feedback link from the legitimate receiver to the legitimate transmitter. The feedback link can be used to transmit a secure key. This secure key can then be used to transmit the message securely over the forward channel via the one-time-pad coding scheme. Hence, an infinite capacity secure feedback channel allows the system to achieve a secrecy capacity equal to the forward channel capacity as if the eavesdropper is not present. It is observed in \cite{Maurer:IT:93}, \cite{Ahlswede:IT:93PI} that even public communication between the legitimate users can enhance the secrecy capacity. It has been shown that positive secrecy capacity can be achieved through public communication even if the eavesdropper's forward channel is \emph{less noisy}. In \cite{Maurer:IT:93} and \cite{Ahlswede:IT:93PI} upper and lower bounds for the perfect secrecy capacity are provided in the case of public communication. These bounds match only for certain special cases. A feedback jamming scheme is also described in \cite{Lai:IT:07} for modulo additive channels.

\begin{figure}[]
\centering
\includegraphics[width=3.25in]{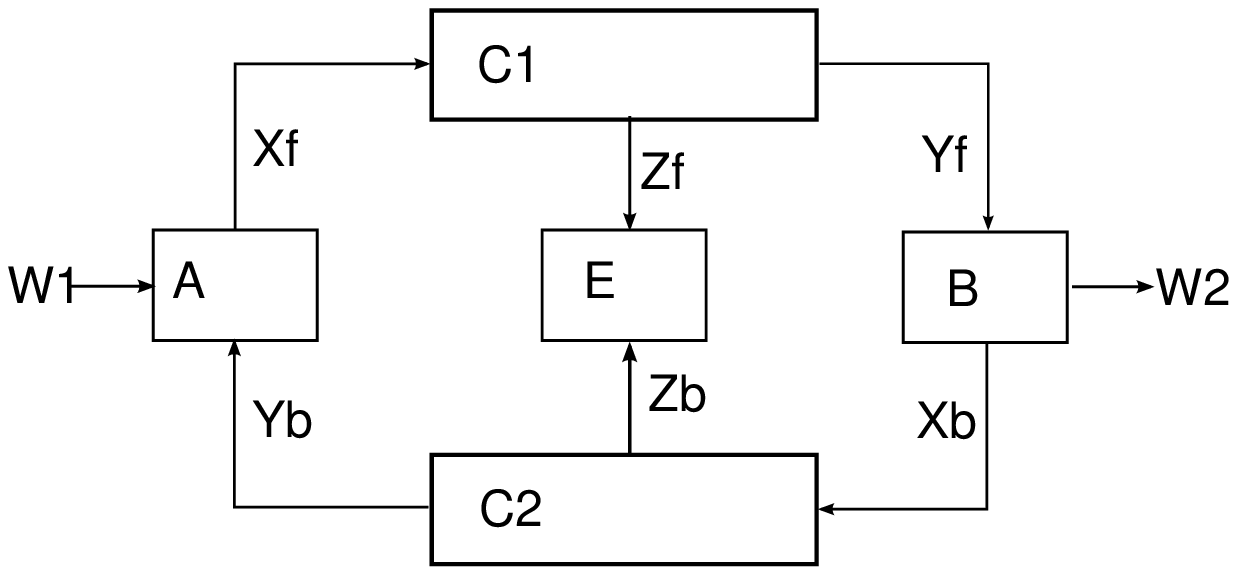}
\caption{Wiretap channel with noisy feedback channel whose output can also be observed by the eavesdropper.}
\label{f:fdbck}
\end{figure}

In this paper, we first propose an achievable secrecy scheme for a general wiretap channel model with feedback (see Fig.~\ref{f:fdbck}), in which the forward channel from Alice to Bob and Eve and the backward channel from Bob to Alice and Eve are orthogonal broadcast channels. The achievability of the proposed scheme follows from using the backward channel for generating a secret key shared by Alice and Bob and then using this secret key  to transmit the message securely over the forward channel via the one-time-pad coding scheme. We then apply this secrecy scheme, in conjunction with Maurer's feedback coding technique \cite{Maurer:IT:93}, to a scenario with  independent binary symmetric forward and backward channels. We explicitly describe the achievable secrecy rates of the proposed scheme and show the improvements in secrecy rate achieved with respect he feedback scheme for binary symmetric channels proposed in \cite{Amariucai:CISS:08}.

\psfrag{W1}{$W$}
\psfrag{W2}{$\hat{W}$}
\psfrag{Xf}{$X_f$}
\psfrag{Yf}{$Y_f$}
\psfrag{Zf}{$Z_f$}
\psfrag{Xb}{$X_b$}
\psfrag{Yb}{$Y_b$}
\psfrag{Zb}{$Z_b$}
\psfrag{C1}[Bl][Bl][0.8]{forward channel}
\psfrag{C2}[Bl][Bl][0.8]{backward channel}
\psfrag{A}{Alice}
\psfrag{B}{Bob}
\psfrag{E}{Eve}

In the second part of this paper, we consider secret communication through a Gaussian wiretap channel with perfect channel output feedback, i.e., Bob's noisy channel output is perfectly available to Alice in a casual manner. This system is consistent with the model considered in the Schalkwijk-Kailath (SK) \cite{Schalkwijk:IT:66} scheme, where perfect causal feedback was shown to greatly simplify the achievability of the point-to-point link from Alice to Bob and also improve the error exponent by exploiting the perfect feedback link in this scenario. The SK scheme is a simple deterministic scheme which is easy to implement and analyze as opposed to Shannon theoretic random coding involving long codewords and high complexity encoding/decoding. We show that, in addition to all these attractive properties, the SK scheme also achieves the secrecy capacity in the presence of an eavesdropper when the eavesdropper (Eve) receives a noisy observation of the feedback from Bob in addition to her own channel output from Alice as depicted in Fig.~\ref{f:SKmodel}.


In the rest of the paper, we use $X^n$ and $X_i^n$ to denote the sequences $(X_1,\ldots,X_n)$ and $(X_i,\ldots,X_n)$, respectively. We also define $[x]^+ = \max\{x, 0\}$.

\section{System Model and Definitions}
\label{sec:system}

In the system shown in Fig.~\ref{f:fdbck}, Alice wants to transmit a message $W \in \mathcal{W}$ to Bob over the discrete memoryless broadcast channel $p(y_f, z_f|x_f)$, where $X_{f} \in \mathcal{X}_f$ is Alice's channel input and $Y_{f} \in \mathcal{Y}_f$ and $Z_{f} \in \mathcal{Z}_f$ are the outputs at Bob and Eve, respectively. There is also an independent feedback channel $p(y_b, z_b|x_b)$, where $X_{b} \in \mathcal{X}_b$ is Bob's feedback channel input and $Y_{b} \in \mathcal{Y}_b$ and $Z_{b} \in \mathcal{Z}_b$ are the outputs at Alice and Eve, respectively. Here, subscripts `$f$' and `$b$' represent \textit{forward} and \textit{backward} channels, respectively.

\begin{defn}
A $(2^{nR}, n)$ code for the above feedback channel is composed of a message $W$ uniformly distributed over set $\mathcal{W}=\{1,\ldots, 2^{nR}\}$, stochastic encoders at Alice $f_i: \mathcal{W} \times \mathcal{Y}_{b}^{i-1} \rightarrow \mathcal{X}_f$ which map the message and the previous feedback outputs to the $i$-th channel input, and stochastic feedback encoders at Bob $g_i: \mathcal{Y}_f^{i-1} \times \mathcal{X}_b^{i-1} \rightarrow \mathcal{X}_b$ which map previous channel outputs and the feedback inputs to the $i$-th feedback input, $i=1,\ldots,n$, and finally a decoder at Bob $h:\mathcal{Y}_f^n \times \mathcal{X}_b^n \rightarrow \mathcal{W}$, which maps the channel outputs and the feedback inputs of Bob to the decoded message $\hat{W}$.
\end{defn}

As usual, the block error probability of a code is defined as
\begin{eqnarray}
P_e^n = \frac{1}{2^{nR}} \sum_{W=1}^{2^{nR}} \mathrm{Pr} \{\hat{W} \neq W\},
\end{eqnarray}
while the equivocation rate is defined as
\begin{eqnarray}
  R^n_e = \frac{1}{n} H(W|Z_f^n, Z_b^n).
\end{eqnarray}

\begin{defn}
A secrecy rate $R$ is achievable if there exists a sequence of $(2^{nR},n)$ codes for which $P_e^n \rightarrow 0$ as $n$ goes to infinity and the equivocation rate satisfies
\[R \leq \lim_{n\rightarrow \infty} R^n_e.\]
\end{defn}

\begin{defn}
The \emph{secrecy capacity} $\mathcal{C}_{sf}$ in the presence of feedback is the highest achievable rate $R$.
\end{defn}

\section{An Achievable Secrecy Rate}\label{s:ach}

In the following theorem, we give a lower bound to the secrecy capacity in the presence of feedback. The achievability follows from using the backward channel for generating a secret key shared by Alice and Bob, and then using this secret key in the forward channel to increase the secrecy capacity of the forward channel.

\begin{thm}\label{t:ach}
Rate $R_s$ is achievable if,
\begin{eqnarray}
0 \leq R_s \leq  \min \{ I(V_f;Y_f), I(V_f;Y_f|U_f) - I(V_f;Z_f |U_f) \nonumber \\
  + I(V_b;Y_b) - I(V_b;Z_b) \}, \nonumber
\end{eqnarray}
for some auxiliary random variables $U_f$, $V_f$, and $V_b$ satisfying $I(U_f;Z_f) \geq I(U_f;Y_f)$ with a joint distribution $p(v_b, x_b,y_b,z_b,u_f,v_f,x_f,y_f,z_f) =p(v_b)$ $\cdot p(x_b|v_b)p(y_b,z_b|x_b)p(u_f)p(v_f|u_f)p(x_f|v_f)p(y_f,z_f|x_f)$, i.e., $U_f \rightarrow V_f \rightarrow X_f \rightarrow (Y_f, Z_f)$ and $V_b \rightarrow X_b \rightarrow (Y_b, Z_b)$.
\end{thm}

\textbf{Proof:} The achievability scheme is based on a separation approach in the sense that, Bob uses the backward channel to generate a shared secret key of rate $R_k$, and then this shared key is used to transmit the message $W$ over the direct channel. We can utilize a block based coding structure, where a secure key is generated in the $i$-th block, $i=1,\ldots,B$, and used in block $i+1$, and the desired rate is achieved in the limit of infinite blocks, i.e., as $B \rightarrow \infty$.

For simplicity, we give the proof for a constant $U_f$. Also, for given $P_{V_fX_fY_fZ_f}$, one can consider an auxiliary channel $P_{X_f|V_f}$  and the code for the induced channel $P_{Y_fZ_f|V_f} = \sum_{x_f} P_{X_f|V_f}(x_f|v_f)P_{Y_fZ_f|X_f}(y_f,z_f|x_f)$. Hence, we prove the achievability of  $\min\{I(X_f;Y_f), I(X_f;Y_f) - I(X_f;Z_f) + I(V_b;Y_b) - I(V_b;Z_b)\}$. The more general proof follows as in \cite{Csiszar:IT:78}. It is possible to generate a secret key $W_k$ at rate \[R_k \triangleq \min\{[I(V_b;Y_b) - I(V_b;Z_b)]^+, I(X_f;Z_f)\}.\] over the backward channel \cite{Csiszar:IT:78}. From the perspective of the forward channel, the problem is now equivalent to finding the secrecy capacity of the broadcast channel with a secret key of rate $R_k$.

Let $R_1 \triangleq I(X_f;Y_f) - I(X_f;Z_f)$. Generate $2^{n I(X_f;Y_f)}$ codewords independent identically distributed (i.i.d.) with probability $p(x_f^n)=\prod_{i=1}^n p(x_{fi})$, and partition these codewords into $2^{nR_1}$ codebooks which we name as $\mathcal{C}_{1}, \ldots, \mathcal{C}_{2^{nR_1}}$. Further divide each subcodebook $\mathcal{C}_i$ into $2^{nR_k}$ smaller codebooks, which are named as $\mathcal{C}_{i,1}, \ldots, \mathcal{C}_{i,2^{nR_k}}$. For each message $w=[w_1, w_2]$, where $w_1 \in [1,2^{nR_1}]$ and $w_2 \in [1, 2^{nR_k}]$, first generate $w_2' = w_2 \oplus w_k~\mathrm{mod} (2^{nR_k})$, and  transmit a codeword chosen uniformly random from the codebook $C_{w_1,w_2'}$. Bob can correctly find $(w_1, w_2')$, hence $w_1$ using the secret key, with high probability for large enough $n$. On the other hand, Eve can determine $w_2'$ and the codeword index within the smallest codebook, but cannot receive any information about $w_1$. Moreover, no information about $w_2$ is revealed to Eve as well, because $w_2'$ is uniformly distributed and independent of $w_2$.

\begin{cor}
If Bob's channel in the forward direction and Alice's channel in the backward direction are both less noisy then Eve's, the highest secrecy rate achievable by the proposed scheme in Theorem \ref{t:ach} can be simplified as
\begin{eqnarray}
R_s &\leq & \min \{ I(X_f;Y_f), I(X_f;Y_f) - I(X_f;Z_f) \nonumber \\
 && ~~~~~~~~~~~~+ I(X_b;Y_b) - I(X_b;Z_b) \}, \nonumber
\end{eqnarray}
for a joint distribution of the form $p(x_b)$ $p(y_b,z_b|x_b)$ $p(x_f)$ $p(y_f,z_f|x_f)$.
\end{cor}

%
%

\section{Secrecy Rates for the Binary Symmetric Wiretap Channel with Feedback}

In this section, we focus on the secrecy rates when both the forward and the backward channels in Fig.~\ref{f:fdbck} are independent binary symmetric channels (BSCs). The system model in this case is fully characterized by the four crossover probabilities $\epsilon_f$, $\delta_f$, $\epsilon_b$, and $\delta_b$, corresponding to the channels Alice$\rightarrow$Bob, Alice$\rightarrow$Eve, Bob$\rightarrow$Alice, and Bob$\rightarrow$Eve, respectively.

This model is also analyzed in \cite{Amariucai:CISS:08} in which a secrecy rate based on the transmission scheme proposed by Maurer in \cite{Maurer:IT:93} is proposed: a random binary sequence $x_b^n$ is transmitted by Bob over the backward channel. Assuming Alice's coded message is $v^n$, she transmits the modulo sum of $v^n$ with the received signal from the backward channel, i.e., $x_f^n = v^n \oplus y_b^n$. If we assume that Alice can transmit $x_f^n$ over a noiseless channel, then Bob can reconstruct $v^n \oplus y_b^n \oplus x_b^n$, while the best Eve can do is to reconstruct $v^n \oplus y_b^n \oplus z_b^n$. This is equivalent to a broadcast channel from Alice to Bob and Eve with cross-over probabilities $\epsilon_b$  and $\epsilon_b+\delta_b-2\epsilon_b\delta_b$, respectively.


We propose here to use a combination of Maurer's scheme with the proposed scheme in Section \ref{s:ach}. The maximum secret key rate that can be generated using the feedback channel is $C_s^b \triangleq [h(\delta_b)-h(\epsilon_b)]^+$. Bob uses the first $\alpha n$ channel uses to generate a secret key of rate $\alpha C_s^b$, where $0 \le \alpha \le 1$ is a design parameter that can be optimized according to the crossover probabilities in order to maximize the total secrecy rate. Bob transmits random bits in the rest of the feedback channel uses.

In the forward channel, we first consider the case $\epsilon_f < \delta_f$. Alice divides the secret message into three parts, all of which are transmitted simultaneously. In the first part, Alice transmits a secret message of rate
\begin{eqnarray*}
  R^s_1 = h(\delta_f)-h(\epsilon_f)
\end{eqnarray*}
over the forward channel using the usual secret coding scheme. Alice can simultaneously transmit a message at rate $1-h(\epsilon_f)-R^s_1 = 1-h(\delta_f)$, which can be received by both Bob and Eve. Alice uses the secure key from the feedback channel as a one-time-pad to transmit securely to Bob at rate
\begin{eqnarray*}
  R^s_2= \min\{1-h(\delta_f), \alpha C_s^b \}.
\end{eqnarray*}
The remaining capacity of the forward channel is then $1-h(\delta_f) - R^s_2 = [1-h(\delta_f)-\alpha C_s^b]^+$. Finally, at this rate, Alice transmits a modulo summed message in the same manner as \cite{Amariucai:CISS:08}, using the random bits received from the second portion of the feedback channel. This transmission occurs at rate
\begin{eqnarray*}
   R^s_3 = & \min\Bigl\{[1-h(\delta_f)-\alpha C_s^b]^+, (1-\alpha)\Bigr\}\\
   &  \cdot (h(\epsilon_b+\delta_b-2\epsilon_b\delta_b)-h(\epsilon_b))
\end{eqnarray*}
In the case when $\epsilon_f \ge \delta_f$, it is impossible to have secret communication without feedback, hence $R^s_1 = 0$. It is straightforward to show in this case that $R^s_2= \min\{1-h(\epsilon_f), \alpha C_s^b \}$ and $R^s_3 =  \min\{[1-h(\epsilon_f)-\alpha C_s^b]^+, (1-\alpha)\} (h(\epsilon_b+\delta_b-2\epsilon_b\delta_b)-h(\epsilon_b))$. In both cases, the total secrecy capacity is then $R^s_1+R^s_2+R^s_3$.

\begin{figure}[]
\centering
\includegraphics[width=3.0in]{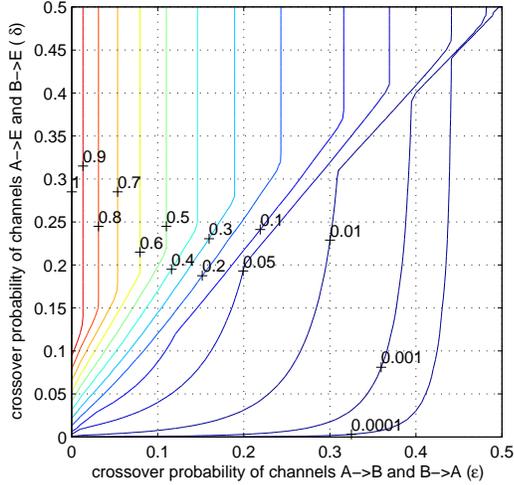}
\caption{Achievable secrecy rate of the proposed feedback scheme with optimized $\alpha$ in binary symmetric channels.}
\label{fig:secrecywithourfeedback}
\end{figure}

To illustrate the gains through the proposed feedback technique, we consider a system model with the same crossover probability in the forward and backward channels, i.e.~$\epsilon =  \epsilon_f =  \epsilon_b$ and $\delta =  \delta_f =  \delta_b$. In Fig.~\ref{fig:secrecywithourfeedback}, we plot the achievable secrecy rate by the proposed transmission scheme with optimized $\alpha$ as a function of $\epsilon$ and $\delta$. As opposed to not having feedback, or using the whole feedback link to generate a secret key, this scheme can achieve positive secrecy rates even in $\delta > \epsilon$ as we have partially incorporated Maurer's coding scheme.  Note also that, our achievable secrecy rates improve upon the ones reported in \cite{Amariucai:CISS:08}.

\begin{figure}[!h]
\centering
\includegraphics[width=3.0in]{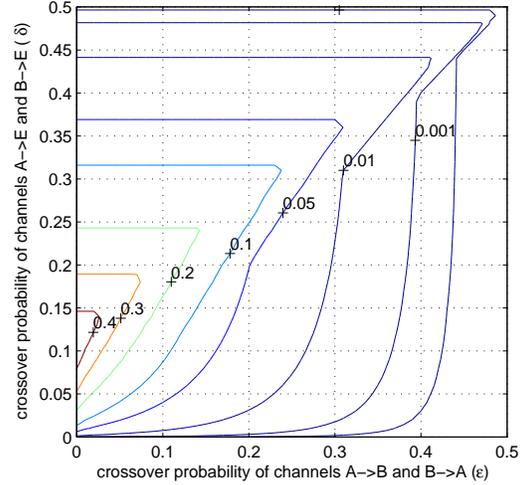}
\caption{Improvement in secrecy rate of the proposed feedback scheme with optimized $\alpha$ in binary symmetric channels with respect to secrecy capacity without feedback.}
\label{fig:secrecyimprovement1}
\end{figure}

Figure~\ref{fig:secrecyimprovement1} plots the improvement in the secrecy rate of the proposed feedback scheme with respect to the secrecy capacity without feedback. Note that the secrecy capacity without feedback is $C_s = [h(\delta)-h(\epsilon)]^+$. Interestingly, the largest gain is obtained at the point $\epsilon = 0$ and $h(\delta) = \frac{1}{2}$, or $\delta \approx 0.11$. At this point the secrecy rate without feedback is $C_s = \frac{1}{2}$, while with feedback we achieve $C_{sf}=1$, an improvement of one half bit. As $\delta \rightarrow \frac{1}{2}$ from this point, the secrecy capacity without feedback increases towards one and feedback results in less improvement.

%


\psfrag{Alice}{Alice}
\psfrag{Bob}{Bob}
\psfrag{Eve}{Eve}
\psfrag{Delay}{Delay}
\psfrag{AN}{$W$}
\psfrag{ANo}{$\hat{W}$}
\psfrag{delta}{$\Delta$}
\psfrag{X}{$X_i$}
\psfrag{Y}{$Y_i$}
\psfrag{Z}{$Z_i$}
\psfrag{M}{$M_i$}
\psfrag{N}{$N_i$}
\psfrag{S}{$S_i$}
\psfrag{Y2}{$Y_{i-1}$}
\psfrag{Ye}{$\bar{Y}_i$}
\psfrag{Ye2}{$\bar{Y}_{i-1}$}

\section{Gaussian Wiretap Channel with Perfect Feedback} \label{s:Gaussian}

\begin{figure}[!h]
\centering
\includegraphics[width=3.2in]{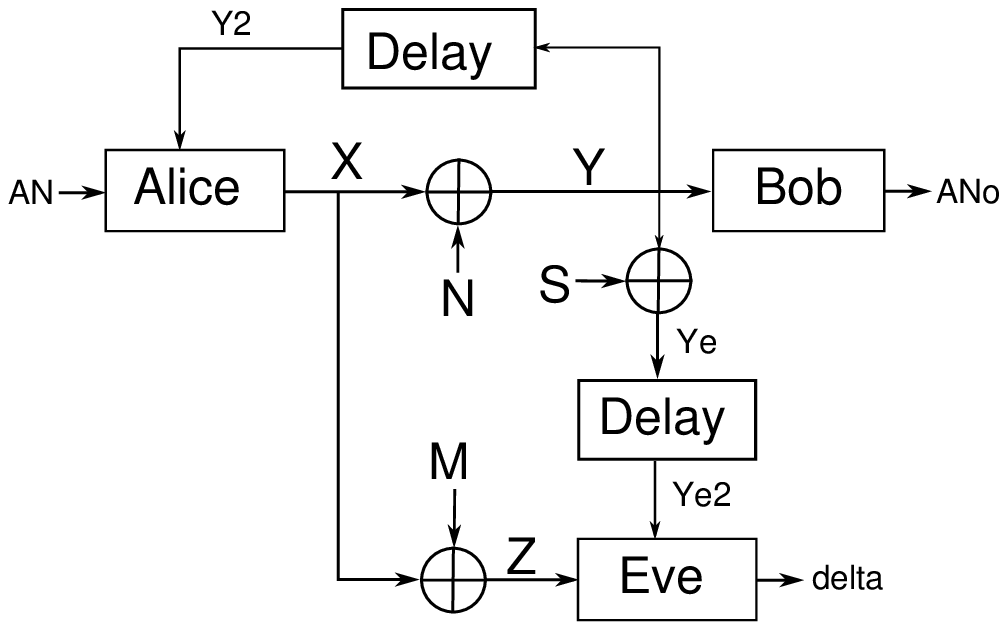}
\caption{The Gaussian wiretap channel with perfect feedback to the legitimate transmitter.}
\label{f:SKmodel}
\end{figure}

Here, we consider a Gaussian wire-tap channel with feedback as seen in Fig.~\ref{f:SKmodel}. The forward channel at time index $i$ is modeled as
\begin{eqnarray}
Y_i = X_i + N_i \mbox{  and } Z_i = X_i + M_i, \nonumber
\end{eqnarray}
in which $N_i$ and $M_i$ are additive white jointly Gaussian noise terms with zero means. There is also an average power constraint $P$ on Alice's transmission. We also have a perfect feedback channel from Bob's output to Alice that operates causally, i.e., at time instant $i$, Alice knows Bob's previous channel outputs $Y^{i-1} = \{Y_1, \ldots, Y_{i-1}\}$. Eve, on the other hand, can only observe a noisy version of this feedback. The feedback from Bob to Alice, as overheard by Eve, is modeled as
\begin{eqnarray}
\bar{Y}_i &=& Y_i + S_i, \label{chmodel3}
\end{eqnarray}
at time $i$, where $S_i$ is also white Gaussian with zero mean. We allow correlation among the additive noise terms of the network at each time instant. The covariance matrix of the noise terms $N, M$ and $S$ is defined as
\begin{eqnarray*}\label{cov}
\mathbf{C} \triangleq \left[ \begin{array}{ccc}
        \sigma_N^2 & \rho_1 \sigma_N\sigma_M & \rho_2 \sigma_N\sigma_S\\
        \rho_1 \sigma_N\sigma_M &  \sigma_M^2 & \rho_3 \sigma_M \sigma_S \\
        \rho_2 \sigma_N\sigma_S & \rho_3 \sigma_M \sigma_S &  \sigma_S^2 \\
        \end{array} \right]
\end{eqnarray*}
which is a real, non-negative definite matrix with $\sigma_M^2>0$, $\sigma_S^2>0$ and $|\rho_i| < 1$, for $i=1,2,3$.

Alice's (potentially stochastic) encoding functions are now defined as $f_i:\mathcal{W} \times \mathcal{Y}^{i-1} \rightarrow \mathcal{X}$. We do not have a channel encoder at Bob, and the perfect feedback scenario is equivalent to having $X_{b,i} = Y_{b,i} = Y_i$ and $Z_{b,i}=\bar{Y}_i$. The average probability of error and the equivocation rate are as defined in Section~\ref{sec:system}.

Ignoring the eavesdropper, the capacity from Alice to Bob (with or without feedback) is $\mathcal{C}_f  = \frac{1}{2}\log\left(1+\frac{P}{\sigma_N^2}\right)$. This also serves as an upper bound on the secrecy capacity with feedback. The secrecy capacity when there is no feedback is given by \cite{Cheong:Stanford:76}
\begin{eqnarray}
\mathcal{C}_s = \frac{1}{2}\left[\log\left(1+\frac{P}{\sigma_N^2}\right) - \log\left(1+\frac{P}{\sigma_M^2}\right)\right]^+.
\end{eqnarray}
Note that $\mathcal{C}_s = 0$ if $\sigma_N^2 \ge \sigma_M^2$.

In \cite{Schalkwijk:IT:66}, Schalkwijk and Kailath proposed a communication scheme (the SK scheme) for a Gaussian channel with perfect feedback that achieves the channel capacity. The SK scheme is based on deterministic coding and parameter estimation. We first give a brief overview of the SK scheme using the notation of \cite{Pillai}.

In the SK scheme, $2^{nR}$ messages are mapped to a signal point by dividing the interval $[-0.5, 0.5]$ into $2^{nR}$ equally spaced subintervals. The mid-point of each subinterval corresponds to a message. Let $\theta$ be the signal point corresponding to the underlying message. At the first transmission, $X_1 = \alpha_1 \theta$ is transmitted, and $Y_1=\alpha_1 \theta + N_1$ is received where $\alpha_1$ is a constant to be chosen. The receiver forms an estimate of $\theta$ based on its observation as
\begin{eqnarray}\label{obsr1}
\hat{\theta}_1 = \hat{X}_1 =\frac{Y_1}{\alpha_1}= \theta + \frac{N_1}{\alpha_1}.
\end{eqnarray}
The transmitter can also compute this estimate using the perfect feedback signal, and in the next transmission, it transmits the estimation error at the receiver, i.e., $X_2 = \alpha_2 (\theta - \hat{\theta}_1) = - \alpha_2 \frac{N_1}{\alpha_1}$, where $\alpha_2$ is another pre-determined constant.
The receiver computes
\begin{eqnarray}\label{obsr2}
\hat{X}_2 = \frac{Y_2}{\alpha_2} +  \frac{Y_1}{\alpha_1} = \theta + \frac{N_2}{\alpha_2}.
\end{eqnarray}
Using the two independent observations of $\theta$ in (\ref{obsr1}) and (\ref{obsr2}), the receiver forms $\hat{\theta}_2$ the maximum likelihood (ML) estimate of $\theta$. Then the transmitter transmits $X_3= \alpha_3 (\theta - \hat{\theta}_2)$, where $\alpha_3$ is another pre-determined constant. Repeating this process, we have
\begin{eqnarray}
X_i &=& \alpha_i (\theta - \hat{\theta}_{i-1}), \label{recur1} \\
\hat{X}_i &=& \hat{\theta}_{i-1} + \frac{Y_i}{\alpha_i}, \mbox{ and }  \label{recur2} \\
\hat{\theta}_i &=& \frac{\sum_{j=1}^{i}\alpha_j^2 \hat{X}_j}{\sum_{j=1}^i \alpha_j^2}  \label{recur3}
\end{eqnarray}
where (\ref{recur3}) is the maximum likelihood (ML) estimate of the parameter $\theta$ at Bob from the observations $Y[1], \dots, Y[i]$.

Now, on choosing $\alpha_i = \gamma \alpha^{i-1}$ with $\gamma = \sqrt{P/\sigma_N^2}$ and $\alpha_1 = \alpha = \sqrt{\frac{P+\sigma_N^2}{\sigma_N^2}}$, it can be shown that the error variance after $n$ iterations is $E[(\theta-\hat{\theta}_n)^2] = \frac{\sigma_N^2}{\alpha^{2n}}$. For $M=2^{nR}$, the probability of error, which corresponds to the probability of $\hat{\theta}_n$ falling outside of the message interval, can be shown to decay to zero exponentially.

In the following theorem, we show that the SK scheme also achieves the optimal secrecy capacity.

\begin{thm}
For the additive white Gaussian noise (AWGN) wire-tap channel with perfect feedback to the transmitter, the secrecy capacity is given by
\begin{eqnarray}
\mathcal{C}_{sf} = \mathcal{C}_f  = \frac{1}{2}\log \left(1+\frac{P}{\sigma_N^2}\right), \label{Csf}
\end{eqnarray}
and this capacity can be achieved by the SK scheme.
\end{thm}

\textbf{Proof:} The converse is obvious since the rate in (\ref{Csf}) is the capacity of the feedback channel without secrecy constraints. Here we prove the achievability of the secrecy capacity in (\ref{Csf}) by the SK scheme.

Fix $\mathcal{W}=\{1, \ldots, 2^{nR}\}$, where $R= \mathcal{C}_{sf}-\epsilon$, for some $\epsilon >0$. Then we know that the average error probability goes to zero as $n \rightarrow \infty$ for any $\epsilon >0$ with the SK scheme.

From the SK scheme, we can observe that
\begin{eqnarray}
\hat{\theta}_i &=& \theta + \frac{\sum_{j=1}^i \alpha_j N_j}{\sum_{j=1}^i \alpha_j^2}, \mbox{ and } \\
X_i &=& - \alpha_i \frac{\sum_{j=1}^{i-1} \alpha_j N_j}{\sum_{j=1}^{i-1} \alpha_j^2} = h_i \sum_{j=1}^{i-1} \alpha_j N_j
\end{eqnarray}
where we have defined $h_i \triangleq   -\frac{\alpha_i}{\sum_{j=1}^{i-1} \alpha_j^2}$.
The observations at Eve are then given as $Z_1 = \alpha_1 \theta + M_1$ and
\begin{eqnarray}
Z_i = \alpha_i \sum_{j=1}^{i-1} h_j N_j + M_i, \mbox{ for } i=2, \ldots, n.
\end{eqnarray}
Finally, the equivocation rate can be written as
\begin{align}
H(\theta|Z_1^n, & \bar{Y}_1^n) \geq  H(\theta|Z_1^n, \bar{Y}_1^n, M_2^n), \label{eqn:1} \\
&=  H(\theta|\alpha_1 \theta + M_1, \alpha_1 \theta + S_1, S_2^n, N^n, M_2^n), \nonumber \\
&=  H(\theta|\alpha_1 \theta + M_1, \alpha_1 \theta + S_1, N_1), \label{eqn:3}
\end{align}
where (\ref{eqn:1}) follows from the fact that conditioning reduces entropy and (\ref{eqn:3}) follows since $N_2^n$ and $M_2^n$ are independent of both $\theta$ and $\alpha_1 \theta + M_1$ due to i.i.d. channel assumption.

The equivocation rate can be further simplified as
\begin{eqnarray*}
H(\theta|Z_1^n, \bar{Y}_1^n) &\geq &  H(\theta|\alpha_1 \theta + M_1, \alpha_1 \theta + S_1, N_1),  \nonumber \\
&=&  H(\theta) - I(\theta ; \alpha_1 \theta + M_1, \alpha_1 \theta + S_1, N_1),  \nonumber \\
&=&  n R - I(\theta ;\alpha_1 \theta + M_1, \alpha_1 \theta + S_1, N_1) ,  \label{eqn:5} \\
&=& nR - I(\theta; \mathbf{A} \theta + \mathbf{B}), \label{eqn:6}
\end{eqnarray*}
where $\mathbf{A} \triangleq [0, \alpha_1, \alpha_1]^T$, $\mathbf{B} \triangleq [N_1, M_1, S_1]^T$,  and where we have used the fact that the message is uniform over the set $\{1, \ldots, 2^{nR}\}$.
The mutual information term in the final expression is difficult to calculate for a uniformly distributed discrete $\theta$, but we know that, allowing for an arbitrary distribution for $\theta$, the mutual information is maximized for a Gaussian input distribution that has the same variance as $\theta$. The variance of $\theta$ as $n$ goes to infinity is $\frac{1}{12}$, hence the corresponding mutual information upper bound is given by
\begin{align}
I(\theta ; \mathbf{A} \theta + \mathbf{B}) &\leq \frac{1}{2} \log \det \left( \mathbf{I} + \frac{1}{12} \mathbf{AA}^T \mathrm{E}[\mathbf{BB}^T]^{-1}  \right) \nonumber \\
&= \frac{1}{2} \log \det \left( \mathbf{I} + \frac{1}{12} \mathbf{AA}^T \mathbf{C}^{-1}\right) \nonumber\\
&= \frac{1}{2} \log \left(1 + \frac{\alpha_1^2 c_1 }{12 \sigma_S^2 \sigma_M^2 c_2}\right)
\end{align}
where
\begin{align}
c_1 &\triangleq 2(\rho_3-\rho_1\rho_2)\sigma_S\sigma_M + (\rho_1^2-1)\sigma_M^2 +(\rho_3^2-1)\sigma_S^2 \mbox{ and }  \nonumber \\
c_2 &\triangleq \rho_1^2 + \rho_2^2 + \rho_3^2 - 2\rho_1\rho_2\rho_3 -1.  \nonumber
\end{align}
Overall, we obtain
\begin{align}
\frac{1}{n} & H(\theta|Z_1^n, \bar{Y}_1^n)  \geq R -  \frac{1}{2n} \log \left(1 + \frac{\alpha_1^2 c_1 }{12 c_2 \sigma_S^2 \sigma_M^2}\right)  \nonumber \\
& = \mathcal{C}_{sf} -\epsilon - \frac{1}{2n} \log \left(1 + \frac{\alpha_1^2 c_1 }{12 c_2\sigma_S^2 \sigma_M^2}\right)  \nonumber \\
& \rightarrow \mathcal{C}_s^f
\end{align}
as $n \rightarrow \infty$ and $\epsilon \rightarrow 0$, if $\sigma_M^2 > 0$, $\sigma_S^2 > 0$ and $c_2 \neq 0$.

Note that, when there is no feedback, the secrecy capacity is nonzero only if $\sigma_N^2 < \sigma_M^2$. However, our result shows that even if the eavesdropper's channel is less noisy than that of the legitimate receiver, the secrecy capacity can be made positive via perfect feedback.

\small
\bibliographystyle{ieeetran}
\bibliography{SKsecrecy}
\end{document}